\documentstyle[12pt]{article}
\global\arraycolsep=2pt
\begin{document}

\begin{titlepage}

\begin{flushright}
CERN-TH.7454/94\\
hep-ph/9409453
\end{flushright}

\vspace{0.5cm}

\begin{center}
\Large\bf Higher-Order Perturbative Corrections\\
to $b\to c$ Transitions at Zero Recoil
\end{center}

\vspace{1.0cm}

\begin{center}
Matthias Neubert\\
{\sl Theory Division, CERN, CH-1211 Geneva 23, Switzerland}
\end{center}

\vspace{1.2cm}

\begin{abstract}
We estimate the two-loop perturbative corrections to zero-recoil
matrix elements of the flavour-changing currents $\bar c\,\gamma^\mu
b$ and $\bar c\,\gamma^\mu\gamma_5\,b$ by calculating the terms of
order $n_f\,\alpha_s^2$ and substituting the dependence on the number
of flavours by the first coefficient of the $\beta$-function. Both
for vector and axial vector currents, we find moderate two-loop
corrections below 1\% in magnitude. Using the
Brodsky--Lepage--Mackenzie prescription to set the scale in the
order-$\alpha_s$ corrections in the $\overline{\rm MS}$ scheme, we
obtain $\mu_V\simeq 0.92\sqrt{m_b m_c}$ and $\mu_A\simeq
0.51\sqrt{m_b m_c}$ in the two cases. These scales are sufficiently
large for perturbation theory to be well-behaved. The implications of
our results to the extraction of $|\,V_{cb}|$ are briefly discussed.
\end{abstract}

\vspace{1.0cm}

\centerline{(Submitted to Physics Letters B)}

\vspace{2.0cm}

\noindent
CERN-TH.7454/94\\
September 1994

\end{titlepage}

\section{Introduction}

It is now widely accepted that the measurement of the $\bar B\to
D^*\ell\,\bar\nu$ decay rate near zero recoil provides for the most
reliable determination of the element $V_{cb}$ of the
Cabibbo--Kobayashi--Maskawa matrix. The theoretical description of
this process has a solid, model-independent foundation based on the
heavy quark expansion, which provides a systematic expansion around
the limit $m_b,m_c\to\infty$. In this limit, QCD exhibits a
spin--flavour symmetry for hadronic systems containing a heavy quark
\cite{Volo}--\cite{review}. The symmetry-breaking corrections are
proportional to powers of $\alpha_s(m_Q)$ or $1/m_Q$, where we use
$m_Q$ as a generic notation for $m_b$ and $m_c$. These corrections
can be investigated in a systematic way using the heavy quark
effective theory (HQET) \cite{Eich}--\cite{Mann}. In particular, at
zero recoil (i.e.\ at equal velocities of the heavy mesons) the
representation of the flavour-changing currents in the HQET reads
\begin{eqnarray}\label{match}
   \bar c\,\gamma^\mu b &\to& \eta_V\,\bar h_v^c\gamma^\mu h_v^b
    + O(1/m_Q^2) \,, \nonumber\\
   \phantom{ \bigg[ }
   \bar c\,\gamma^\mu\gamma_5\,b &\to& \eta_A\,\bar h_v^c\gamma^\mu
    \gamma_5\,h_v^b + O(1/m_Q^2) \,,
\end{eqnarray}
where $h_v^Q$ are the velocity-dependent heavy quark spinors of the
HQET. Hadronic matrix elements of the effective current operators are
normalized because of heavy quark symmetry. The coefficients $\eta_V$
and $\eta_A$ in (\ref{match}) take into account finite
renormalizations of the currents in the intermediate region
$m_b>\mu>m_c$. They can be obtained from an on-shell matching of
current matrix elements in QCD with the corresponding matrix elements
in the HQET. From a measurement of the $\bar B\to D^*\ell\,\bar\nu$
decay rate near zero recoil, one can extract the product
$|\,V_{cb}|\,\eta_A\,(1+\delta_{1/m^2})$, where $\delta_{1/m^2}$
stands for non-perturbative power corrections of order $(\Lambda_{\rm
QCD}/m_Q)^2$ \cite{review,Vcb}.

As very precise experimental data on this decay mode become available
\cite{Ritch}, a detailed theoretical analysis of the
symmetry-breaking corrections to the heavy quark limit becomes
increasingly important. The power corrections $\delta_{1/m^2}$ have
recently been the subject of intense interest
\cite{FaNe}--\cite{Vcbnew}. Here we shall focus on the perturbative
coefficient $\eta_A$ and its analogue for the vector current,
$\eta_V$. At the one-loop order, these coefficients have been known
for a long time \cite{Pasc}:
\begin{eqnarray}\label{oneloop}
   \eta_V &=& 1 + C_F\,{\alpha_s\over 4\pi}\,\phi(z) \,,
    \nonumber\\
   \eta_A &=& 1 + C_F\,{\alpha_s\over 4\pi}\,
    \big[ \phi(z) - 2 \big] \,,
\end{eqnarray}
where $C_F=4/3$ is a colour factor, $z=m_c/m_b$, and
\begin{equation}
   \phi(z) = - 3\,{1+z\over 1-z}\,\ln z - 6
   = {\ln^2\!z\over 2} - {\ln^4\!z\over 120} + {\ln^6\!z\over 5040}
   + O(\ln^8\!z) \,,
\end{equation}
with $\phi(1)=0$. Using $\mu=\sqrt{m_b m_c}$ for the scale in the
running coupling constant, one obtains $\eta_V\simeq 1.02$ and
$\eta_A\simeq 0.97$. Throughout this work, we use the input
parameters $m_b=4.80$~GeV, $m_c=1.44$ GeV, and $\Lambda_{\rm
QCD}=0.11$ GeV in the one-loop expression for the running coupling
constant in the $\overline{\rm MS}$ scheme (for $n_f=4$). This gives
$\alpha_s(m_b)\simeq 0.20$, $\alpha_s(m_c)\simeq 0.29$, and
$\alpha_s(\sqrt{m_b m_c})\simeq 0.24$. The fact that the one-loop
corrections are smaller than the naive expectation of
$\alpha_s/\pi\sim 10\%$ makes one suspicious about the importance of
higher-order corrections. A renormalization-group improvement of
(\ref{oneloop}) has been performed, which sums logarithms of the type
$(\alpha_s\ln z)^n$, $\alpha_s(\alpha_s\ln z)^n$ and $z(\alpha_s\ln
z)^n$ to all orders in perturbation theory \cite{Falk}--\cite{QCD}.
However, since in the case of $b\to c$ transitions $\ln z$ is not a
particularly large parameter, one expects that the residual two-loop
corrections not included in this procedure are as important as some
of the logarithmic terms. Therefore, a complete two-loop calculation
seems worth while. In particular, it would help to reduce the scale
ambiguity in the above one-loop results. Unfortunately, however, such
a calculation appears to be rather tedious for the two-scale problem
at hand. In this letter, we derive partial results for the two-loop
corrections to $\eta_V$ and $\eta_A$, which may be used to obtain an
estimate of the size of the full corrections. Moreover, our analysis
will allow us to study the convergence of perturbation theory for
$b\to c$ transitions. It is thus of interest even beyond the two-loop
order. We emphasize that our main goal is to investigate whether
there are indications for large higher-order terms in the
perturbative series for $\eta_V$ and $\eta_A$, and not so much to
obtain predictions for these quantities that are more precise than
existing ones. To this end, it would be necessary to perform the
complete two-loop calculations.

Let us write the perturbative series for any one of the coefficients
$\eta_V$ and $\eta_A$ in the form
\begin{equation}\label{etasum}
   \eta - 1 = \sum_{n=1}^\infty \bigg(
   {\alpha_s(\mu)\over 4\pi} \bigg)^n\,\eta_n(\mu) \,,
\end{equation}
where $\alpha_s(\mu)$ is the running coupling constant renormalized
at some scale $\mu$. Since $\eta$ is renormalization-group invariant,
the $\mu$-dependence on the right-hand side must cancel between the
expansion coefficients and the running coupling constant. It is
useful to make explicit the dependence of the coefficients
$\eta_n(\mu)$ on the number of quark flavours. In the case at hand,
the first dependence on $n_f$ comes at the two-loop order from
diagrams containing a quark loop in a gluon propagator. In general,
we may write
\begin{equation}
   \eta_n(\mu) = c_{n,n-1}(\mu)\,\beta_0^{n-1}
   + c_{n,n-2}(\mu)\,\beta_0^{n-2} + \ldots
   + c_{n,0}(\mu) \,;\quad n\ge 1 \,,
\end{equation}
where $\beta_0=11 - \frac{2}{3}\,n_f$ is the first coefficient of the
$\beta$-function. In particular, at the two-loop order we have
\begin{equation}\label{eta}
   \eta - 1 = {\alpha_s(\mu)\over 4\pi}\,c_{1,0}(\mu)
   + \bigg( {\alpha_s(\mu)\over 4\pi}\bigg)^2\,\Big[
   c_{2,1}(\mu)\,\beta_0 + c_{2,0}(\mu) \Big] + \ldots \,.
\end{equation}
For the case of currents composed of one heavy and one light quark,
it has been found in explicit calculations that there are large
two-loop coefficients when the scale in the running coupling constant
is chosen to be the heavy quark mass $m_Q$, and that these large
coefficients are dominated by the term proportional to
$c_{2,1}\,\beta_0$ in (\ref{eta}) \cite{BGnew}. This empirical
observation can be understood if one assumes that in these on-shell
calculations the relevant scale is much below the ``natural'' scale
$m_Q$, meaning that in loop diagrams virtual momenta below $m_Q$ give
a sizeable contribution. Because of the relation
\begin{equation}\label{rescale}
   \alpha_s(\kappa\mu) = \alpha_s(\mu)\,\sum_{n=0}^\infty
   \bigg( {\beta_0\,\alpha_s(\mu)\over 4\pi} \bigg)^n\,
   \big( -\ln\kappa^2 \big)^n + \ldots \,,
\end{equation}
using an inadequate scale can induce large higher-order coefficients
$c_{n,n-1}$. It is possible to absorb some of these large corrections
by using a lower scale. However, in some cases this scale turns out
to be too low for perturbation theory to be well-defined. Some of the
heavy--light currents considered in Ref.~\cite{BGnew} provide an
example of this phenomenon. As we will discuss below, another
example is provided by inclusive decays of hadrons containing a heavy
quark \cite{LSW}. One of our purposes here is to investigate if
something similar happens for currents composed of two heavy quarks.

\section{Large-$n_f$ asymptotics of perturbation theory}

The above discussion justifies that a calculation of the coefficients
$c_{n,n-1}$, and in particular of the two-loop coefficient $c_{2,1}$,
is worth while. Not only can it serve as an estimate of the size of
the full two-loop correction, but also to choose an appropriate
scale in the order-$\alpha_s$ term. Technically, the coefficients
$c_{n,n-1}$ can be projected out by considering the formal limit of
large $n_f$, in which the series (\ref{etasum}) takes the form
\begin{eqnarray}
   \eta - 1 &=& {1\over\beta_0}\,\sum_{n=1}^\infty \bigg(
    {\beta_0\,\alpha_s(\mu)\over 4\pi} \bigg)^n\,c_{n,n-1}(\mu)
    + O(1/n_f^2) \nonumber\\
   &=& {1\over\beta_0}\,\sum_{n=1}^\infty\,
    \ln^{-n}\!\big( \mu^2/\Lambda_{\rm QCD}^2 \big)\,c_{n,n-1}(\mu)
    + O(1/n_f^2)\,.
\end{eqnarray}
Note that $\beta_0$ is of order $n_f$, whereas the product
$\beta_0\,\alpha_s$ is of order $n_f^0$. A convenient way to analyse
this series is by considering its Borel transform with respect to
$\ln(\mu^2/\Lambda_{\rm QCD}^2)$ \cite{tHof}, which we define
as\footnote{This definition differs from the one adopted in
Ref.~\cite{Chris} by a factor $1/\beta_0$.}
\begin{equation}\label{Fexp}
   \widetilde F(u,\mu) = \sum_{n=1}^\infty
   {u^{n-1}\over\Gamma(n)}\,c_{n,n-1}(\mu) \,.
\end{equation}
The function $\widetilde F(u,\mu)$ can be calculated using the
renormalon calculus of Beneke and Braun \cite{Bene,BBren}. In
Ref.~\cite{Chris}, this technique has been used to calculate the
Wilson coefficients of flavour-changing heavy quark currents at
arbitrary velocity transfer. It is straightforward to specialize the
results to zero recoil to obtain explicit expressions for the Borel
transforms of the coefficients $\eta_V$ and $\eta_A$. We find
\begin{eqnarray}
   \widetilde F_{V,A}(u,\mu) &=& C_F\,e^{-C u}\,\bigg(
    {\mu^2\over m_b m_c} \bigg)^u\,
    {\Gamma(u)\,\Gamma(1-2u)\over\Gamma(2-u)}\,\Bigg\{
    {2(1\pm u)\over 2-u}\,{z^u - z^{1-u}\over 1-z} \nonumber\\
   &&\qquad\mbox{}+ {2(1-u)\over 1+2u}\,
    {z^{-u} - z^{1+u}\over 1-z} + {1+z\over 1-z}\,
    \big( z^{-u} - z^u \big) \Bigg\} \\
   &&\mbox{}- 3 C_F\,e^{-C u}\,\bigg[
    \bigg( {\mu\over m_b} \bigg)^{2u} + \bigg(
    {\mu\over m_c} \bigg)^{2u} \bigg]\,(1-u^2)\,
    {\Gamma(u)\,\Gamma(1-2 u)\over\Gamma(3-u)} \,, \nonumber
\end{eqnarray}
where again $z=m_c/m_b$. The upper (lower) sign in the first term in
parenthesis refers to the vector (axial vector) current. $C$ is a
scheme-dependent constant, with $C=-5/3$ in the $\overline{\rm MS}$
scheme. The scheme- and scale-dependence of $\widetilde F(u,\mu)$
cancels against the scheme- and scale-dependence of the running
coupling constant when one inverts the Borel transformation using the
integral relation
\begin{equation}
   \eta - 1 = {1\over\beta_0}\,\int\limits_0^\infty\!{\rm d}u\,
   \bigg( {\Lambda_{\rm QCD}^2\over\mu^2} \bigg)^u\,
   \widetilde F(u,\mu) + O(1/n_f^2) \,,
\end{equation}
since the product
\begin{equation}\label{LamC}
   {\Lambda_{\rm QCD}^2\over\mu^2}\,e^{-C}\,\mu^2
   = \Lambda_{\rm QCD}^2\,e^{-C}
   = \Lambda_{\overline{\rm MS}}^2\,e^{5/3}
\end{equation}
is scheme- and scale-independent.

According to (\ref{Fexp}), the coefficients $c_{n,n-1}$ can be
obtained from a expansion of the Borel transform in powers of $u$.
Substituting the result back into (\ref{eta}), we find that the
$\mu$-dependence indeed cancels. At the two-loop order, we obtain
\begin{eqnarray}
   \eta_V &=& 1 + {\bar\alpha_s\over 4\pi}\,C_F\,\phi(z)
    + \bigg( {\bar\alpha_s\over 4\pi} \bigg)^2\,\Big[
    C_F\,\Big( -C-{\textstyle\frac{3}{2}} \Big)\,\phi(z)\,\beta_0
    + c_{2,0}^V(z) \Big] + \ldots \,, \nonumber\\
   \eta_A &=& 1 + {\bar\alpha_s\over 4\pi}\,C_F\,
    \Big[ \phi(z) - 2 \Big] \\
   &&\phantom{ 1 }
    + \bigg( {\bar\alpha_s\over 4\pi}
    \bigg)^2\,\bigg\{ C_F\,\Big( -C\,\Big[ \phi(z) - 2 \Big]
    - {\textstyle\frac{5}{6}}\,\phi(z) +1 \Big)\,\beta_0
    + c_{2,0}^A(z) \bigg\} + \ldots \,, \nonumber
\end{eqnarray}
where $\bar\alpha_s\equiv\alpha_s(\sqrt{m_b m_c})$. Since by charge
conservation the vector current is not renormalized for $z=1$, it
follows that $c_{2,0}^V(1)=0$. For $z=m_c/m_b=0.3$, we obtain in the
$\overline{\rm MS}$ scheme (with $C=-5/3$)
\begin{eqnarray}
   \eta_V &\simeq& 1 + 0.236\,{\bar\alpha_s\over\pi}
    + \Big( 0.082 + {\textstyle\frac{1}{16}}\,c_{2,0}^V(z) \Big)\,
    \bigg({\bar\alpha_s\over\pi}\bigg)^2 + \ldots  \,,
    \nonumber\\
   \eta_A &\simeq& 1 - 0.431\,{\bar\alpha_s\over\pi}
    + \Big( -1.211 + {\textstyle\frac{1}{16}}\,c_{2,0}^A(z) \Big)\,
    \bigg({\bar\alpha_s\over\pi}\bigg)^2 + \ldots  \,,
\end{eqnarray}
where we use $\beta_0=25/3$, corresponding to $n_f=4$, which is
appropriate for the intermediate region $m_b>\mu>m_c$. The partial
two-loop corrections that we have computed amount to very moderate
effects, which however have the same sign as the one-loop
corrections. Numerically, with $\bar\alpha_s\simeq 0.24$, we obtain
$\delta\eta_V\simeq 5\times 10^{-4}$ for the corresponding
contribution to $\eta_V$, and $\delta\eta_A\simeq -7\times 10^{-3}$
for the contribution to $\eta_A$. Thus, we find no indication for
large two-loop corrections in the case of heavy--heavy currents. This
is in stark contrast to the case of heavy--light currents, where the
coefficient of the $(\alpha_s/\pi)^2$ term is typically of order 10,
with $\frac{1}{16}\,c_{2,0}$ of order unity \cite{BGnew}.

\section{BLM scale setting}

Brodsky, Lepage and Mackenzie (BLM) have advocated to absorb vacuum
polarization effects into the running coupling constant by choosing
the scale in the order-$\alpha_s$ correction so that there are no
corrections of order $\beta_0\,\alpha_s^2$ in an expansion such as
(\ref{eta}) \cite{BLM}. This physically appealing scale-setting
prescription usually results in a reasonable perturbative series.
Accepting this point of view, one may argue that perturbation theory
works well in cases where the BLM scale is sufficiently large,
whereas it breaks down if this scale is too low. Recently, it has
been shown that the BLM scale to be used in {\it inclusive\/} $\bar
B\to X\,\ell\,\bar\nu$ decays is very low, $\mu_{\rm incl} \simeq
0.07\,m_b$, indicating a breakdown of perturbation theory \cite{LSW}.
In fact, it had been noted before that the one-loop corrections to
the inclusive decay rate exhibit a strong scale dependence
\cite{BaNi}. This observation puts severe limitations on the
usefulness of inclusive decays for the determination of $|\,V_{cb}|$.
At least, a calculation of the two-loop corrections is necessary
before a reliable analysis can be performed. Fortunately, as we will
now show, the situation appears to be much better for the {\it
exclusive\/} decay $\bar B\to D^*\ell\,\bar\nu$.

{}From (\ref{eta}) and (\ref{rescale}), it follows that for a general
perturbative series the BLM scale is given by
\begin{equation}
   \mu_{\rm BLM} = \exp\Bigg( - {c_{2,1}(\mu)\over 2 c_{1,0}(\mu)}
   \Bigg)\,\mu \,,
\end{equation}
which can be shown to be $\mu$-independent. Note that the BLM scale
is not scheme-indepen\-dent; instead, it is such that the value of
$\alpha_s(\mu_{\rm BLM})$ is scheme-independent. According to
(\ref{LamC}), this requires that $\mu_{\rm BLM}\propto e^{C/2}$.
Indeed, from our calculation in the previous section, we obtain
\begin{eqnarray}
   \mu_V &=& e^{3/4}\,e^{C/2}\,\sqrt{m_b m_c} \,, \nonumber\\
   \phantom{ \Bigg[ }
   \mu_A &=& \exp\Bigg\{ {6-5\phi(z)\over 12 (2-\phi(z))}\bigg\}\,
    e^{C/2}\,\sqrt{m_b m_c} \,.
\end{eqnarray}
For $z=0.3$ and in the $\overline{\rm MS}$ scheme (with $C=-5/3$),
this yields $\mu_V\simeq 0.920\,\sqrt{m_b m_c}$ and $\mu_A\simeq
0.509\,\sqrt{m_b m_c}$. These scales are large enough for
perturbation theory to be well-behaved. The corresponding
scheme-independent values of the running coupling constant are
$\alpha_s(\mu_V)\simeq 0.24$ and $\alpha_s(\mu_A) \simeq 0.30$. Using
these coupling constants instead of $\alpha_s(\sqrt{m_b m_c})$ in the
one-loop expressions (\ref{oneloop}) changes the values of $\eta_V$
and $\eta_A$ by $\delta\eta_V\simeq 5\times 10^{-4}$ and
$\delta\eta_A\simeq -9\times 10^{-3}$. These changes are practically
identical to our estimates of the two-loop corrections at the end of
the previous section.

\section{Conclusions}

We have presented a partial calculation of the two-loop matching
corrections to the flavour-changing currents in $b\to c$ transitions
at zero recoil. Both for vector and axial vector currents we find
small corrections, which are below 1\% in magnitude. Using the
Brodsky--Lepage--Mackenzie prescription to set the scale in the
order-$\alpha_s$ corrections, we obtain $\alpha_s(\mu_V) \simeq 0.24$
and $\alpha_s(\mu_A)\simeq 0.30$ for the relevant coupling constants
in the two cases. The fact that the corresponding scales are
sufficiently large indicates good convergence of perturbation theory
for {\it exclusive\/} $b\to c$ transitions. This is in contrast to
{\it inclusive\/} $B$ decays, where it was found that the appropriate
scale is as low as 350 MeV, indicating a breakdown of perturbation
theory \cite{LSW}. These findings support that the exclusive
semileptonic decay $\bar B\to D^*\ell\,\bar\nu$ is the
``gold-plated'' mode for a precision measurement of $|\,V_{cb}|$
\cite{Vcbnew}.

It is interesting to compare our estimate of the size of the two-loop
corrections with the intrinsic uncertainty in $\eta_V$ and $\eta_A$,
which results from the necessity to regularize the divergent
asymptotic behaviour of the perturbative series for these quantities.
In fact, both $\eta_V$ and $\eta_A$ are known to contain infrared
renormalons, which lead to ambiguities of order $(\Lambda_{\rm
QCD}/m_Q)^2$. A measure of the resulting intrinsic uncertainty is
\cite{Chris}
\begin{equation}
   \Delta\eta_{V,A} = {3\beta_0\over 32}\,\bigg[ \Delta m\,
   \bigg( {1\over m_c} \mp {1\over m_b} \bigg) \bigg]^2 \,,
\end{equation}
where $\Delta m\sim\Lambda_{\rm QCD}$ is the renormalon ambiguity in
the pole mass of a heavy quark \cite{BBren,Bigiren}. As previously,
the upper (lower) sign refers to the vector (axial vector) current.
Assuming $\Delta m\simeq 0.1$ GeV, we obtain $\Delta\eta_V\simeq
0.2\%$ and $\Delta\eta_A\simeq 0.6\%$. These numbers match nicely
with our estimate of the two-loop corrections, indicating that it is
sufficient and adequate to truncate the perturbative series at the
two-loop order.

In conclusion, we thus believe that the existing calculations of the
matching corrections for heavy quark currents, for instance the
one-loop results $\eta_V\simeq 1.02$ and $\eta_A\simeq 0.97$ obtained
from (\ref{oneloop}), or the values $\eta_V=1.025\pm 0.015$ and
$\eta_A=0.985\pm 0.015$ obtained by performing a next-to-leading
order renormalization-group improvement \cite{QCD}, are reliable at
the level of a few per cent. There are no indications for unusually
large higher-order corrections.

\bigskip
{\it Acknowledgements:\/}
It is a pleasure to thank Thomas Mannel for useful discussions and
Chris Sachrajda for collaboration on subjects closely related to this
work. I am indebted to David Broadhurst and Andrey Grozin for making
the results of Ref.~\cite{BGnew} available to me prior to
publication.

\end{document}